\documentclass{PoS}

\title{Search for First Generation Leptoquarks in ep Collisions at HERA}

\ShortTitle{Search for First Generation Leptoquarks in ep Collisions at HERA}

\author{\speaker{David M. South}\thanks{on behalf of the H1 Collaboration.}\\
  Deutsches Elektronen Synchrotron\\
  Notkestrasse 85\\
  22607 Hamburg, Germany\\
  E-mail: \email{david.south@desy.de}}

\abstract{
A search for first generation scalar and vector leptoquarks produced in $ep$ collisions is
performed by the H1 Collaboration at HERA.
The full H1 data sample is used in the analysis, corresponding to an integrated luminosity
of $446$~pb$^{-1}$.
No evidence for the production of leptoquarks is observed in final states with a large
transverse momentum electron or with large missing transverse momentum, and
constraints on leptoquark models are derived.
For leptoquark couplings of electromagnetic strength $\lambda=0.3$, first generation
leptoquarks with masses up to $800$~GeV are excluded at $95\%$ confidence level.
}

\FullConference{36th International Conference on High Energy Physics,\\
		July 4-11, 2012\\
		Melbourne, Australia}

\begin{document}

\section{Leptoquark production at HERA}
\label{sec:theory}

The $ep$ collisions at HERA provide a unique possibility to search for 
new particles coupling directly to a lepton and a quark.
Leptoquarks (LQs), colour triplet bosons that do just that, are an example
of such particles and appear in many theories attempting to unify the
quark and lepton sectors of the Standard Model (SM).


A discussion of the phenomenology of LQs at HERA can be found
elsewhere~\cite{Adloff:1999tp}.
In the framework of the Buchm\"uller-R\"uckl-Wyler (BRW) effective model~\cite{BRW},
LQs are classified into $14$ types with respect to the quantum numbers spin $J$,
weak isospin $I$ and chirality $C$, resulting in seven scalar ($J=0$) and seven vector ($J=1$) LQs.
Whereas all $14$ LQs couple to electron\footnote{Here the term ``electron'' is used generically to
refer to both electrons and positrons, unless otherwise stated.}-quark pairs, four of the left-handed LQs,
namely $S_0^L$, $S_{1}^L$, $V_0^L$ and $V_{1}^L$, may also decay to a neutrino-quark pair.
In particular, for $S_0^L$ and $V_0^L$ the branching fraction of decays into an
electron-quark pair is predicted by the model to be
\mbox{$\beta_{e}\!=\!\Gamma_{\rm eq}/(\Gamma_{\rm eq}+\Gamma_{\rm \nu_{e} q})\!=0.5$},
where \mbox{$\Gamma_{\rm eq}$} (\mbox{$\Gamma_{\rm \nu_{e} q}$}) denotes the partial
width for the LQ decay to an electron (neutrino) and a quark $q$.
The branching fraction of decays into a neutrino-quark pair is then given by
$\beta_{\nu_{e}} =1-\beta_{e}$.


Leptoquarks carry both lepton ($L$) and baryon ($B$) quantum numbers, and the
fermion number \mbox{$F\!=\!L\!+\!3\,B$} is assumed to be conserved.
Leptoquark processes at HERA proceed directly via $s$-channel resonant LQ production
or indirectly via $u$-channel virtual LQ exchange.
A dimensionless parameter $\lambda$ defines the coupling at the
lepton-quark-LQ vertex.
For LQ masses well below the centre-of-mass energy $\sqrt{s}=319$~GeV,
the $s$-channel production of $F = 2$ ($F = 0$) LQs in $e^-p$ ($e^+p$) collisions dominates.
However, for LQ masses above $\sqrt{s}$, both the $s$ and $u$-channel processes are important
such that both $e^-p$ and $e^+p$ collisions have similar sensitivity to all LQs types.


The analysis presented here examines LQ decays to a quark and a first generation lepton,
following the flavour conservation implicit in the BRW model.
In a more general extension of this analysis, dedicated searches have been performed by H1
for second and third generation leptoquarks, examining final states containing a quark and
a charged lepton of a different flavour, i.e. a muon or tau lepton~\cite{Aktas:2007ji,h1lfv2011}.
A search for second and third generation leptoquarks was recently performed using the
full H1 $ep$ data sample taken at $\sqrt{s} = 319$~GeV, where
for a coupling strength of $\lambda = \sqrt{4 \pi \alpha_{\rm em}} =0.3$, LQs
decaying with the same coupling strength to a muon-quark pair or a tau-quark pair are
excluded at $95\%$ confidence level (CL) up to leptoquark masses of $712$~GeV and
$479$~GeV, respectively~\cite{h1lfv2011,ichep2012lfv}.

\section{Search for first generation leptoquarks}

This search considers final states where the leptoquark decays into an electron and a quark
$ep \rightarrow eX$ or a neutrino and a quark $ep \rightarrow \nu_{e} X$.
The full H1 $ep$ data sample has now been analysed~\cite{h1lq2011}, which
comprises $164$~pb$^{-1}$ recorded in $e^{-}p$ collisions and $282$~pb$^{-1}$ in $e^{+}p$
collisions, of which $35$~pb$^{-1}$ were recorded at $\sqrt{s} = 301$~GeV.
Data collected from 2003 onwards were taken with a longitudinally polarised lepton beam.
As leptoquarks are chiral particles, these data are analysed in four separate polarisation samples,
formed by combining data periods with similar lepton beam charge $e^{\pm}$ and polarisation
$P_{e} = (N_{R} - N_{L})/(N_{R} + N_{L})$, where $N_{R}$ ($N_{L}$) is the number of right (left)
handed leptons in the beam.


First generation LQ decays lead to topologies similar to those of deep-inelastic scattering
(DIS) neutral current (NC) and charged current (CC) interactions at high negative
four-momentum transfer squared $Q^2$.
The analysis is therefore performed using DIS event selections similar to those used in
inclusive DIS analyses~\cite{Aaron:2012qi} and previous first generation
LQ searches~\cite{Aktas:2005pr}.
%
%
Neutral current events are selected by requiring a scattered electron with energy
$E_{e'} > 11$~GeV and $Q^{2}~>~133$~GeV$^2$ and in the inelasticity region
$0.1~<~y<~0.9$.
Background from neutral hadrons or photons misidentified as leptons is suppressed
by requiring a charged track to be associated to the lepton candidate.
%
%
Charged current events are selected by requiring significant missing transverse
momentum $P_{T}^{\rm miss}>12$~GeV, which is due to the undetected neutrino,
in the inelasticity region $0.1 <  y < 0.85$.
Photoproduction background is suppressed by exploiting the correlation
between $P_{T}^{\rm miss}$ and the ratio $V_{\rm ap} / V_{\rm p}$ of transverse energy flow
anti-parallel and parallel to the hadronic final state transverse momentum
vector~\cite{Adloff:1999tp}.
Further details of the event selection can be found in the H1 publication~\cite{h1lq2011}.

\begin{figure}[ht] 
  \begin{center}
   \includegraphics[width=0.85\textwidth]{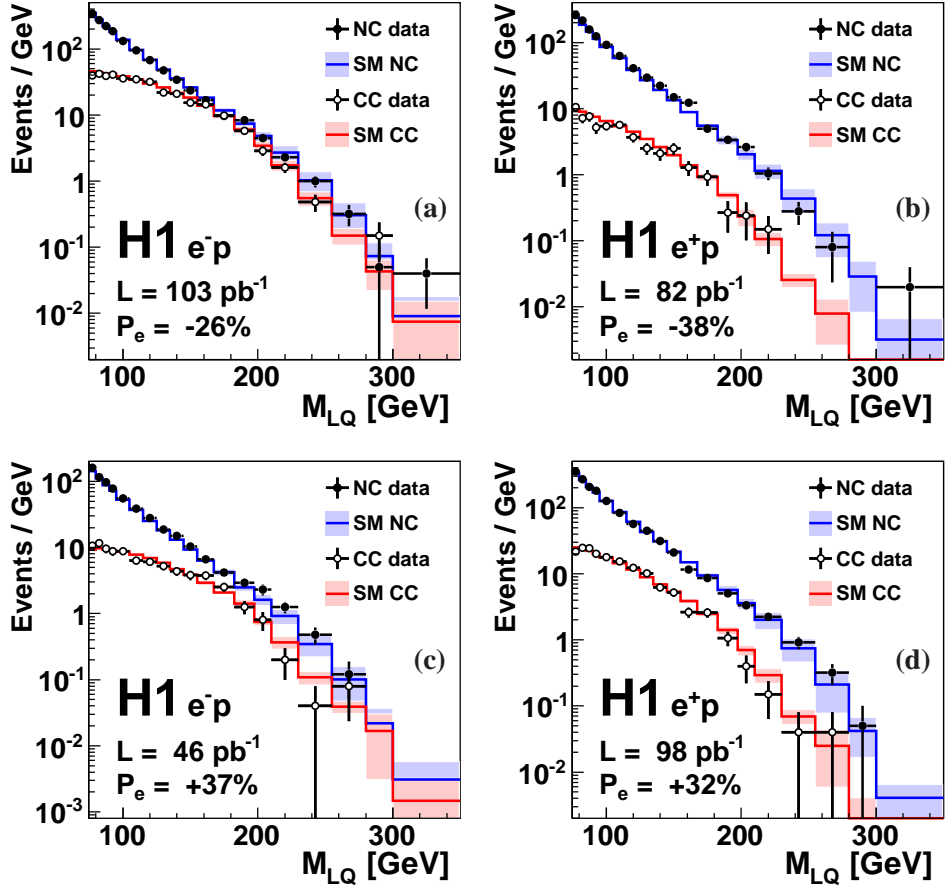}     
  \end{center} 
  \begin{picture} (0.,0.)
  \setlength{\unitlength}{1.0cm}
    \put (6.75,10){\bf\normalsize (a)}
    \put (13.15,10){\bf\normalsize (b)}
    \put (6.75, 4){\bf\normalsize (c)}
    \put (13.15,4){\bf\normalsize (d)}
 \end{picture}
 \vspace{-1cm}
   \caption{The reconstructed leptoquark mass from the search for first generation leptoquarks
      using the 2003-2007 H1 data, which was taken with a longitudinally polarised lepton beam. The left-handed
      electron data (a) and left-handed positron data (b) are shown in the top row; the right-handed electron data (c) and
      right-handed positron data (d) are shown in the bottom row.
      The luminosity and average longitudinal lepton polarisation of each data set is indicated.
      The NC (solid points) and CC (open points) data are compared to the SM predictions
      (histograms), where the shaded bands indicate the total SM uncertainties.}
 \label{fig:polmassplots}
\end{figure}

A good description of the H1 data by the SM is observed, where the expectation is dominated
by DIS processes in all event samples, with small additional contributions from photoproduction.
Mass spectra of the four H1 data sets taken with a longitudinally polarised lepton
beam are shown in figure~\ref{fig:polmassplots}, where both the NC and CC event samples
are presented.
Since no evidence for LQ production is observed in any of the NC or CC data samples,
the data are used to set constraints on leptoquarks coupling to first generation fermions.

\section{First generation leptoquark limits}

In the absence of a signal, the results of the search are interpreted in terms of
exclusion limits on the mass and the LQ coupling.
The data are studied in bins in the $M_{LQ}-y$ plane, where the NC and CC data samples
with different lepton beam charge and polarisation are kept as distinct data sets.
Limits are determined from a statistical analysis which uses the method of fractional event 
counting, optimised for the presence of systematic uncertainties.
A frequentist analysis is performed of a test statistic obtained from the data for each
leptoquark type, mass and coupling hypothesis.
A full description of the statistical analysis and limit procedure employed can be
found elsewhere~\cite{h1lq2011}.

\begin{figure}[] 
\begin{center}
      \includegraphics[width=.49\textwidth]{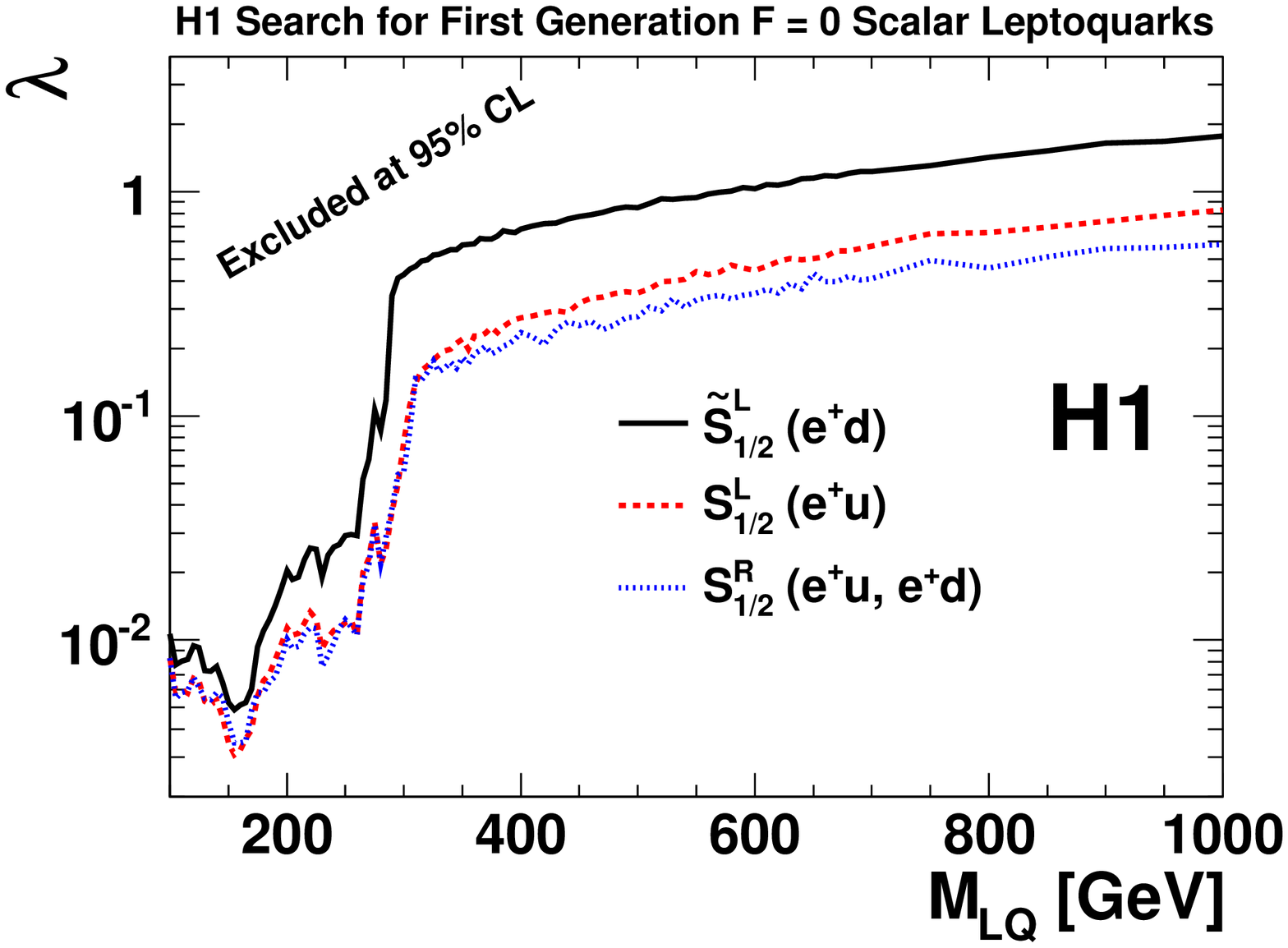}
      \includegraphics[width=.49\textwidth]{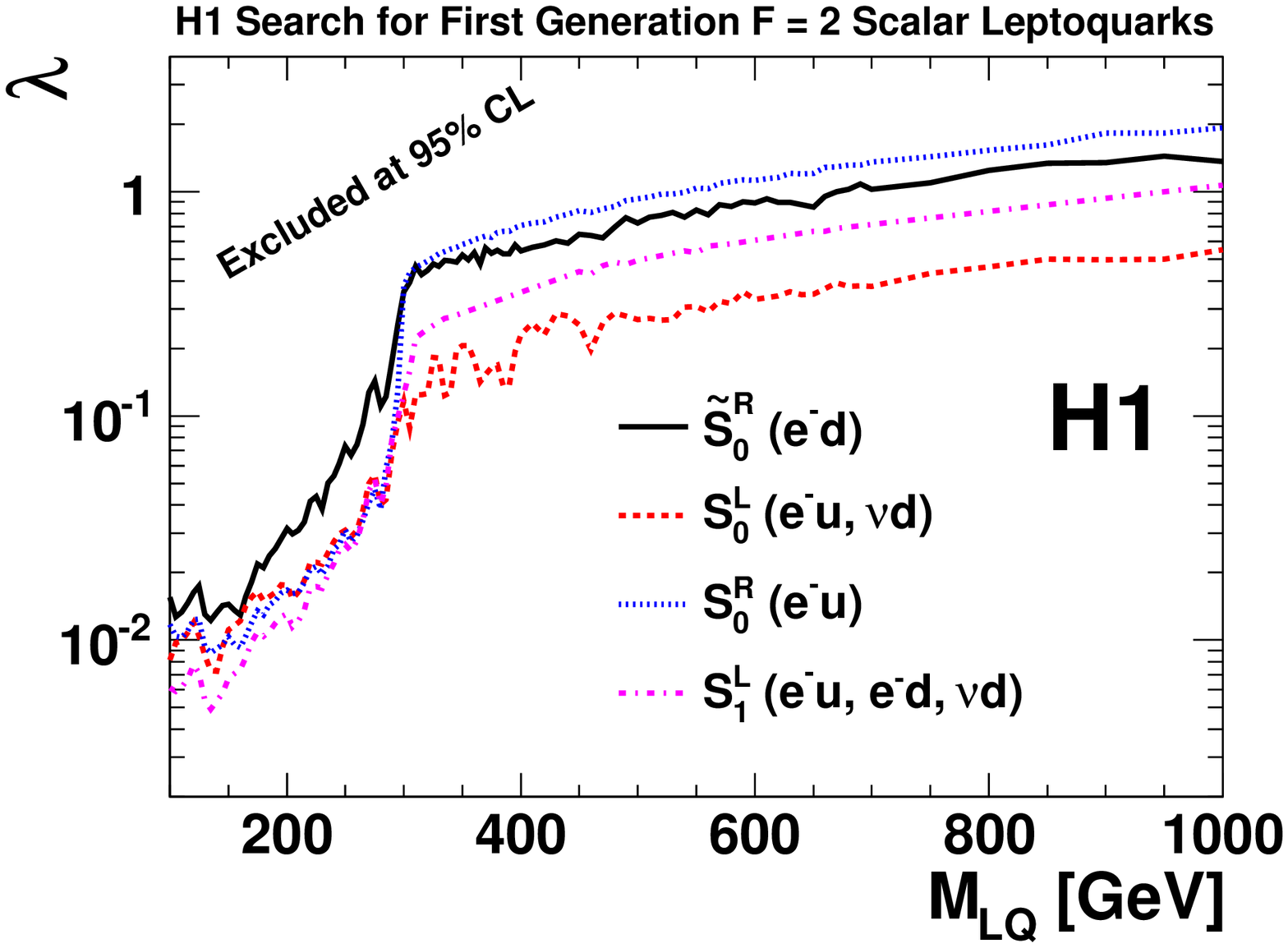}
      \includegraphics[width=.49\textwidth]{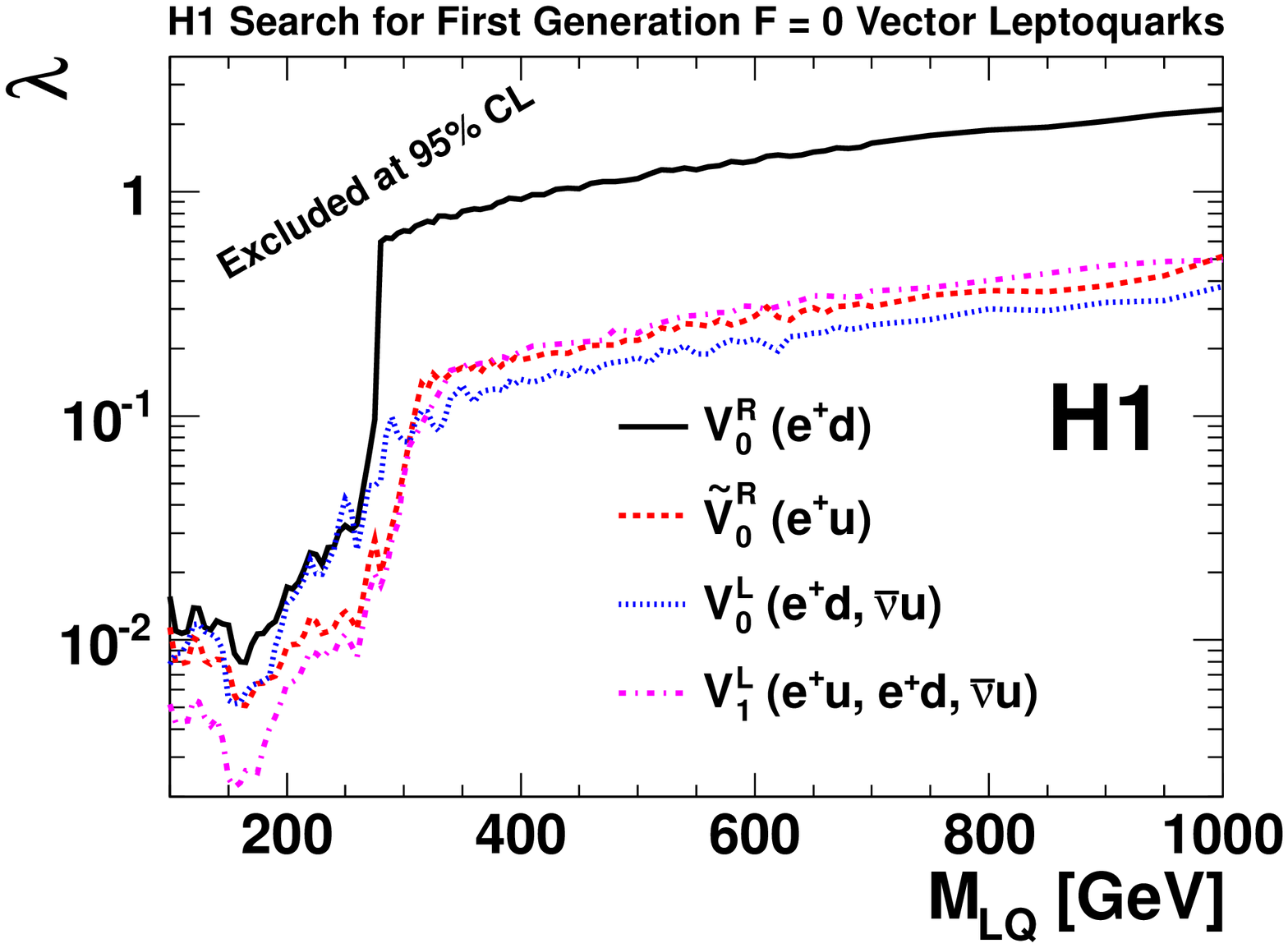}
      \includegraphics[width=.49\textwidth]{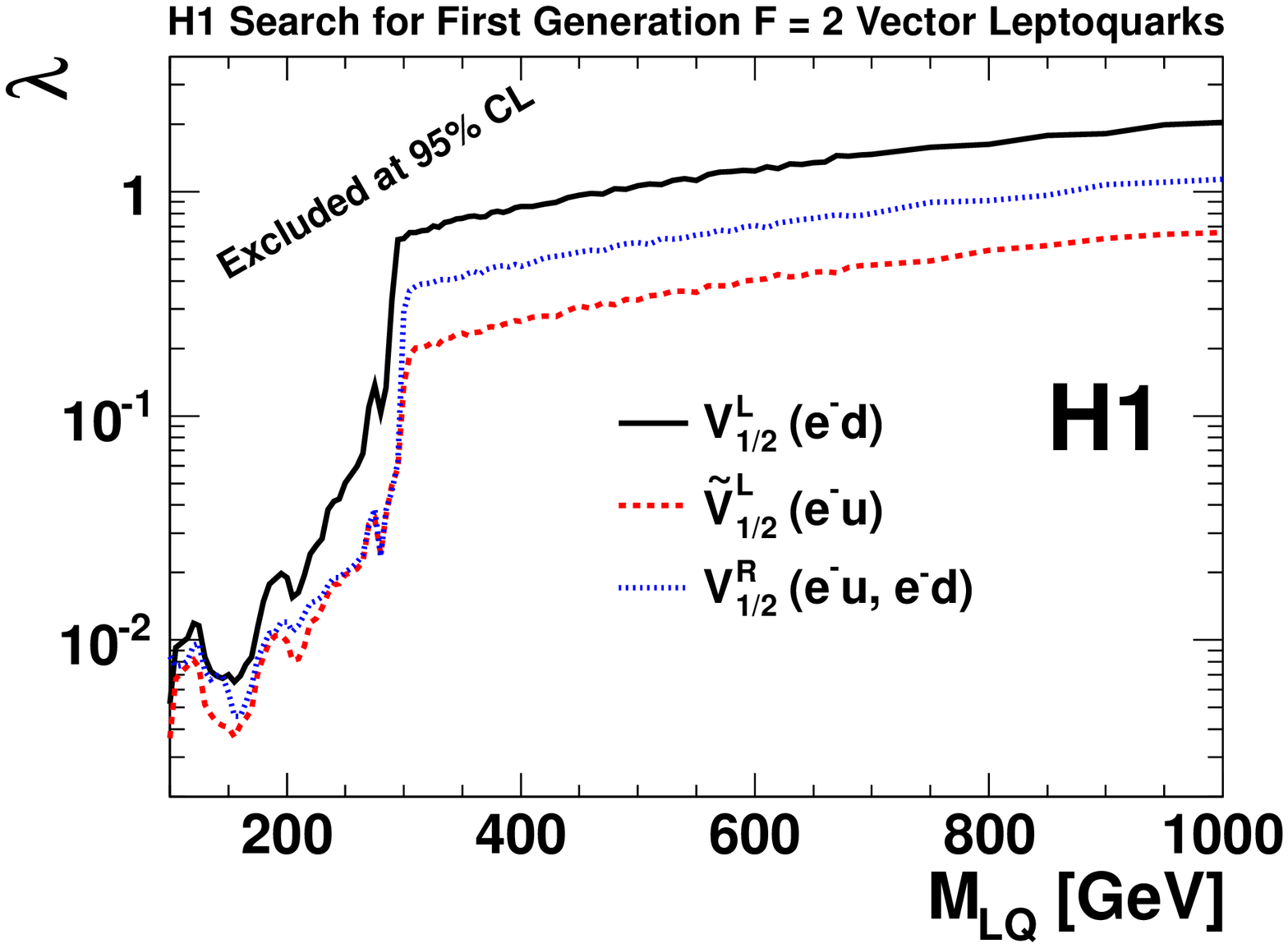}
\end{center}
  \begin{picture} (0.,0.)
  \setlength{\unitlength}{1.0cm}
    \put (2.9,8.2){\bf\normalsize (a)}
    \put (10.5,8.2){\bf\normalsize (b)}
    \put (2.9,2.6){\bf\normalsize (c)}
    \put (10.5,2.6){\bf\normalsize (d)}
 \end{picture}
\vspace{-1cm}
  \caption{Exclusion limits for the 14 leptoquarks (LQs) described by the Buchm\"uller, R\"uckl and
    Wyler model. The limits are expressed on the coupling $\lambda$ as a function of leptoquark
    mass for the scalar LQs with (a) $F = 0$ and (b) $F = 2$ and the vector LQs with (c) $F = 0$ and (d) $F = 2$.
    Domains above the curves are excluded at $95\%\,{\rm CL}$. The parentheses after the LQ name indicate the
    fermion pairs coupling to the LQ, where pairs involving anti-quarks are not shown.}
\label{fig:limits}
 \end{figure} 

Upper limits on the coupling $\lambda$ obtained at $95$\%~CL are shown as a function of the
leptoquark mass in figure~\ref{fig:limits}, displayed as groups of scalar and vector LQs for
both $F = 2$ and $F = 0$.
For LQ masses near the kinematic limit of $319$~GeV, the limit corresponding
to a resonantly  produced LQ turns smoothly into a limit on the virtual effects of both an
off-shell $s$-channel LQ process and a $u$-channel LQ exchange.
For LQ masses much greater than the HERA centre-of-mass energy the two processes
contract to an effective four-fermion interaction.
For a coupling of electromagnetic strength $\lambda = \sqrt{4 \pi \alpha_{\rm em}} = 0.3$,
LQs produced in $ep$ collisions decaying to an electron-quark or a neutrino-quark pair are
excluded at $95\%$~CL up to leptoquark masses between $277$~GeV~($V_{0}^{R}$)
and $800$~GeV~($V_{0}^{L}$), depending on the leptoquark type.

\begin{figure}[] 
  \begin{center}
    \includegraphics[width=0.85\textwidth]{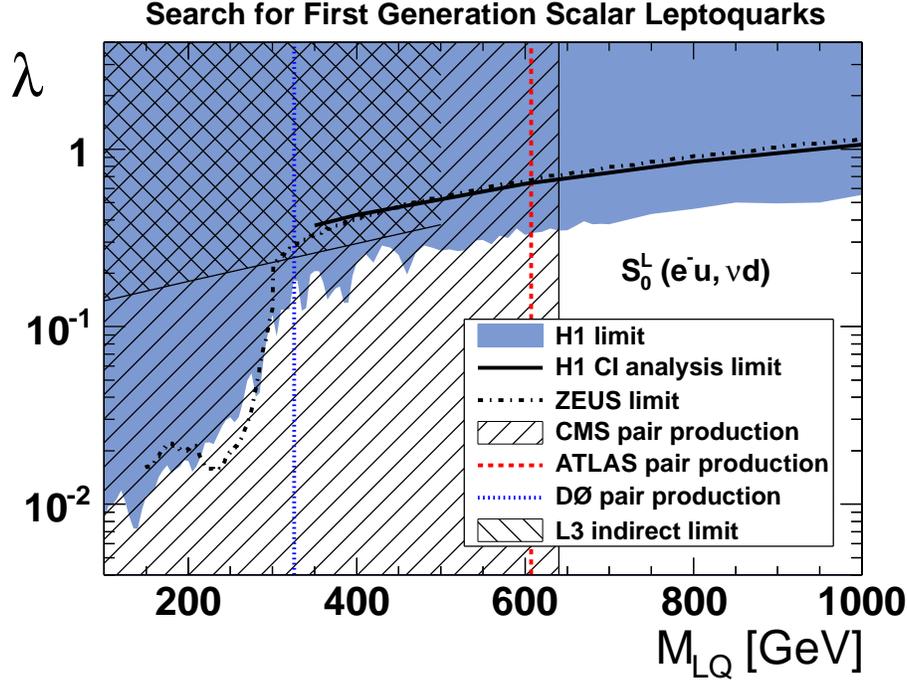}     
  \end{center} 
  \vspace{-0.5cm}
  \caption{Exclusion limits in the framework of the Buchm\"uller, R\"uckl and
    Wyler model on the coupling $\lambda$ as a function of leptoquark mass for the $S_{0}^{L}$
    leptoquark, which has a branching fraction $\beta_{e} = 0.5$.
    Domains above the curves and to the left of the vertical lines are excluded at $95$\%~CL.
    Limits from the D{\O}, L3 and ZEUS experiments and those from
    the LHC (CMS and ATLAS, $\sqrt{s} =7$ TeV data) are shown for comparison, as well as constraints
    on LQs with masses above $350$~GeV from the H1 contact interaction (CI) analysis.}
  \label{fig:compare}
\end{figure}

Within the framework of the BRW model, the $S_{0}^{L}$ LQ decays to both an electron-quark pair
and a neutrino-quark pair, resulting in $\beta_{e} = 0.5$.
The H1 limits on $S_{0}^{L}$ are compared to those from other experiments
in figure~\ref{fig:compare}, including the similar limit from the ZEUS experiment~\cite{Abramowicz:2012tg}.
The indirect limit from a search for new physics in $e^+e^-$ collisions at LEP by the
L3 experiment~\cite{Acciarri:2000uh} is also indicated, as well as $\beta_{e} = 0.5$ limits
from D{\O}~\cite{Abazov:2009gf} at the Tevatron and from ATLAS~\cite{Aad:2011ch} and
CMS~\cite{CMS:2012dn} at the LHC, based on $\sqrt{s}=7$~TeV data taken in 2011.
The H1 limits at high leptoquark mass values are also compared to those
obtained in a dedicated contact interaction analysis~\cite{Aaron:2011mv}.
The additional impact of the CC data can be seen, where a stronger limit is achieved in the
LQ analysis, compared to the contact interaction analysis which is based only on NC data.
The limits from hadron colliders are based on searches for LQ pair-production and are
independent of the coupling $\lambda$, where the strongest current limit for
$\beta_{e} = 1.0$ ($\beta_{e} = 0.5$) scalar LQs is $830$~GeV ($640$~GeV)
as reported by the CMS collaboration.
For a leptoquark mass of $640$~GeV, this analysis rules out the $S_{0}^{L}$ LQ
for coupling strengths larger than about $0.35$.

\newpage


\begin{thebibliography}{9}

\bibitem{Adloff:1999tp}
  C.~Adloff {\it et al.}  [H1 Collaboration],
  {\it A search for leptoquark bosons and lepton flavour violation in $e^{+}p$ collisions at HERA},
  {\it Eur.\ Phys.\ J.\ C} {\bf 11} (1999) 447
  [Erratum-ibid.\ {\bf 14} (2000) 553]
  [{\tt hep-ex/9907002}].

\bibitem{BRW}  W.~Buchm\"uller, R.~R\"uckl and D.~Wyler,
 {\it Leptoquarks in lepton-quark collisions},
 {\it Phys.\ Lett.\ B} {\bf 191} (1987) 442  [Erratum-ibid. {\bf 448} (1999) 320].

\bibitem{Aktas:2007ji}
  A.~Aktas {\it et al.}  [H1 Collaboration],
  {\it Search for lepton flavour violation in $ep$ collisions at HERA},
  {\it Eur.\ Phys.\ J.\ C} {\bf 52} (2007) 833
  [{\tt hep-ex/0703004}].

 \bibitem{h1lfv2011}
  F.~D.~Aaron {\it et al.} [H1 Collaboration],
  {\it Search for lepton flavour violation at HERA},
  {\it Phys.\ Lett.\ B} {\bf 701 } (2011)  20
  [{\tt arXiv:1103.4938}].

\bibitem{ichep2012lfv}
  D.~M.~South,
  {\it Search for lepton flavour violation at HERA},
  in proceedings of {\it 36th International Conference on High Energy Physics}, {\tt PoS(ICHEP2012)140}.


\bibitem{h1lq2011}
  F.~D.~Aaron {\it et al.} [H1 Collaboration],
  {\it Search for first generation leptoquarks in $ep$ collisions at HERA},
  {\it Phys.\ Lett.\ B} {\bf 704} (2011) 338 
  [{\tt arXiv:1107.3716}].

\bibitem{Aaron:2012qi}
  F.~D.~Aaron {\it et al.}  [H1 Collaboration],
  {\it Inclusive deep inelastic scattering at high $Q^2$ with longitudinally polarised lepton beams at HERA},
  {\it JHEP} {\bf 1209} (2012) 061
  [{\tt arXiv:1206.7007}].

\bibitem{Aktas:2005pr}
  A.~Aktas {\it et al.}  [H1 Collaboration],
  {\it Search for leptoquark bosons in $ep$ collisions at HERA},
  {\it Phys.\ Lett.\ B} {\bf 629} (2005) 9
  [{\tt hep-ex/0506044}].


\bibitem{Abramowicz:2012tg}
  H.~Abramowicz {\it et al.}  [ZEUS Collaboration],
  {\it Search for first-generation leptoquarks at HERA},
  {\it Phys.\ Rev.\ D} {\bf 86} (2012) 012005
  [{\tt arXiv:1205.5179}].

\bibitem{Acciarri:2000uh}
  M.~Acciarri {\it et al.} [L3 Collaboration],
  {\it Search for manifestations of new physics in fermion pair production at LEP},
  {\it Phys.\ Lett.\ B} {\bf 489} (2000)  81
  [{\tt hep-ex/0005028}].

\bibitem{Abazov:2009gf}
  V.~M.~Abazov {\it et al.}  [D{\O} Collaboration],
  {\it Search for pair production of first-generation leptoquarks in $p\bar{p}$ collisions at $\sqrt{s}  = 1.96$~TeV},
  {\it Phys.\ Lett.\  B} {\bf 681} (2009) 224
  [{\tt arXiv:0907.1048}].

\bibitem{Aad:2011ch}
  G.~Aad {\it et al.}  [ATLAS Collaboration],
  {\it Search for first generation scalar leptoquarks in $pp$ collisions at $\sqrt{s}=7$ TeV with the ATLAS detector},
  {\it Phys.\ Lett.\ B} {\bf 709} (2012) 158
  [Erratum-ibid.\  {\bf 711} (2012) 442]
  [{\tt arXiv:1112.4828}].

\bibitem{CMS:2012dn}
  S.~Chatrchyan {\it et al.}  [CMS Collaboration],
  {\it Search for pair production of first- and second-generation scalar leptoquarks in $pp$ collisions at $\sqrt{s}= 7$ TeV},
  {\it Phys.\ Rev.\ D} {\bf 86} (2012) 052013
  [{\tt arXiv:1207.5406}].

\bibitem{Aaron:2011mv}
  F.~D.~Aaron {\it et al.}  [H1 Collaboration],
  {\it Search for contact interactions in $e^{\pm}p$ collisions at HERA},
  {\it Phys.\ Lett.\ B} {\bf 705} (2011) 52
  [{\tt arXiv:1107.2478}].

\end{thebibliography}
\end{document}